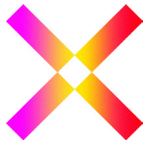

# Patterns in Deep Time

**Dave Griffiths**
dave@thentrythis.org
Then Try This, Penryn, England

**Elizabeth Wilson**
ewilson01@protonmail.ch
Queen Mary, University of London, England

**Iván Paz**
ivan@toplap.cat
TOPLAP Barcelona, Spain

**Alex McLean**
alex@slab.org
Then Try This, Sheffield, England

*With the collaboration of:*
**Joana Chicau**
**Flor de Fuego**
**Timo Hoogland**
**Eloi Isern**
**Michael-Jon Mizra**
**Robert Pibernat**

In this paper, we explore how textile pattern-making can be a useful activity for live coders used to manipulating software. We ran an algorithmic patterns workshop in July 2022 — with a node at on_the_fly. collect(_) festival in Barcelona, a node in Sheffield and the workshop leader in Penryn — where we created an activity recreating ancient patterns by weaving on tablet looms that we constructed from card and yarn, and sent to the participants for this remote/multi location workshop. One of the aims of the Algorithmic Pattern project is to highlight the relationship people have had with patterns over history, and how we can use this to uncover certain misconceptions we have about algorithmic patterns in contemporary society. We collected responses from those who participated in the workshop and collated the responses. We found that tablet weaving allows us to connect the physical patterns with their algorithmic descriptions. Also, errors relate with the trade-off among expectations and surprise and exploring new unexpected possibilities. Finally, sharing the experiences among the participants allows us to observe how we interpret patterns when comparing it with other experiences.

# 1. Introduction

This paper introduces a workshop exploring algorithmic patterns (Mclean 2020) across both textiles and the contemporary practice of live coding (Blackwell et al. 2022). First, we share background to the thinking that led up to this workshop, providing both recent and historical cultural reference points. Then we will share experiences and reflections as workshop participants and conclude with further reflections on where this line of research will end up next.

Through the workshop, we explored algorithms in terms of their physical movements and outcomes, their representations, their ability to convey meaning, and how the act of weaving connects and reflects such dimensions. Moreover, comparing digital computers with handweaving revealed the textile-like patterning of binary operations, which is not always obvious when algorithms are expressed in high level programming languages and abstractions. On a more general level this paper reflects on computation as a human activity that extends beyond automation through electronic computers. The workshop did not aim for prescriptive goals and outcomes, but rather was cast as an exploratory activity that attempted to compare and contrast the practices of those used to working with algorithms as source code, with heritage algorithms (Eglash et al. 2019) in textiles.

Both handweaving and programming involve the exploration of entangled, countable, discrete structures. Both also involve the authorship and manipulation of digital representations, such as grid-based block designs and drawdowns in the case of weaving, and source code in programming. They are also both digital in their physical manifestations — e.g., the high/low voltages which manifest computer processes, and up/down interlacements which form weaves. However, while weaving and programming are both digital crafts (McLean, Fanfani, and Harlizius-Klück 2018), weaving is considerably older, having developed over thousands of years. This is important to remember, and key to the motivation for this workshop. Our contention is that as the far older, and more highly developed human digital artform, both culturally and technologically, programmers have a great deal to learn from weaving.

The workshop itself took place in a networked hub format in July 2022, split over three locations — with a node at the on_the_fly.collect(_) festival based at the Hangar.org space in Barcelona, a node hosted by the Then Try This studio in Sheffield, and the workshop leader based in the main Then Try This studio in Penryn, Cornwall. The workshop was convened by Iván Paz and Lizzie (Elizabeth) Wilson in Barcelona, and Alex McLean in Sheffield, with the support of the rest of the on-the-fly project team including Ludovica Michelin and Lina Bautista. The other named authors contributed reflections to this paper as workshop participants.

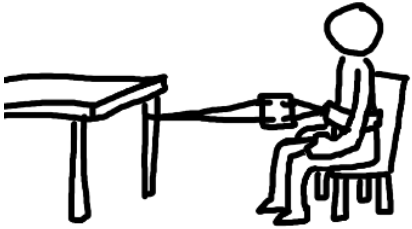

**Figure 1.1:** A tablet weaver showing warp threads tensioned using a backstrap and passing through the tablets to create a 'shed' gap for passing the weft.

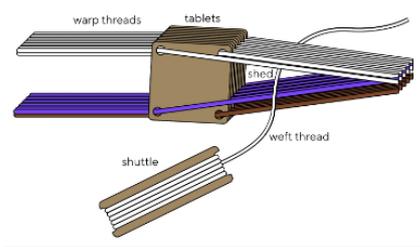

**Figure 1.2:** Diagram showing each warp thread passing through one of four holes in the tablets, creating a shed through which the weft is passed.

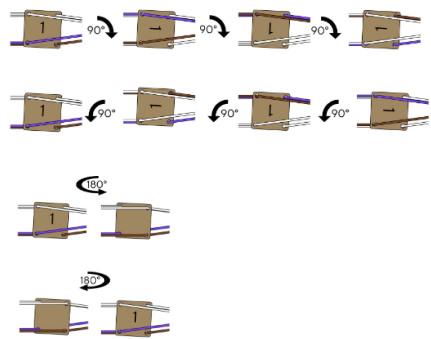

**Figure 1.3:** Diagram showing the different sheds created by a) turning and b) flipping the tablets. One or more such turning and flipping actions may be performed, potentially to different groups of tablets, between the passing of each weft thread.

## 2. Tablet Weaving

The particular form of weaving introduced in the workshop was *tablet weaving,* also known as *card weaving.* Like all textile crafts, tablet weaving is structurally (and culturally) distinct, for example expertise and knowledge of knitting or even other weaving techniques does not naturally transfer to tablet weaving. Indeed, it would take more than a lifetime to explore all the complexities of tablet weaving. Although tablet weaving shares fundamental elements with other forms of weaving such as tensioned warp threads and passing of perpendicular weft threads, the addition of twisting and flipping actions make it closer to braiding or twining in some respects. A detailed introduction to the complexities of tablet weaving is out of scope of the present paper, and fully grasping its nature requires hands-on practice, but figures 1.1-1.3 convey the fundamental elements.

Although the principles of tablet weaving are relatively straightforward, the patterns that emerge can be difficult to comprehend. The turning and flipping of cards interacts with previous states in the weave, creating three-dimensional interference patterns as the differently coloured threads twist into view. Learning tablet weaving is therefore full of surprise, with complex patterns emerging from simple movement sequences. A common experience is to puzzle over how such a pattern appears on the front of the woven band, only to find a completely different pattern on the reverse side of the band. In the end, the beginner settles into a mode of experimentation, trying out different patterns to get a tacit feel for the potential outcome. This mode of experimentation with algorithm and material has parallels with the experience of live coding; a primary motivation for bringing the two practices together.

## 3. Digital vs. Physical in Weaving and Programming

The workshop was led by Dave Griffiths, based on their personal experience of learning weaving and programming simultaneously when very young. This early foundation provided a certain way of seeing the parallels between these digital worlds — not in analogy or on the surface level but providing two views on the same underlying cosmos. Indeed, an additional core motivation for the workshop was to consider how the digital and the physical have become separated in contemporary culture. We set the scene for this by discussing a meeting between the UK prime minister and advisors when rapidly deciding policy in the days leading up to the first COVID-19 lockdown. In a space devoid of internet access, except one smartphone (belonging to Dominic Cummings, the Chief Adviser to the British Prime Minister at the time), they were desperately trying to understand all the issues involved on the single white board shown in Fig. 2.

**Figure 2:** The whiteboard used to plan the UK government's initial covid response on 13 March 2020.

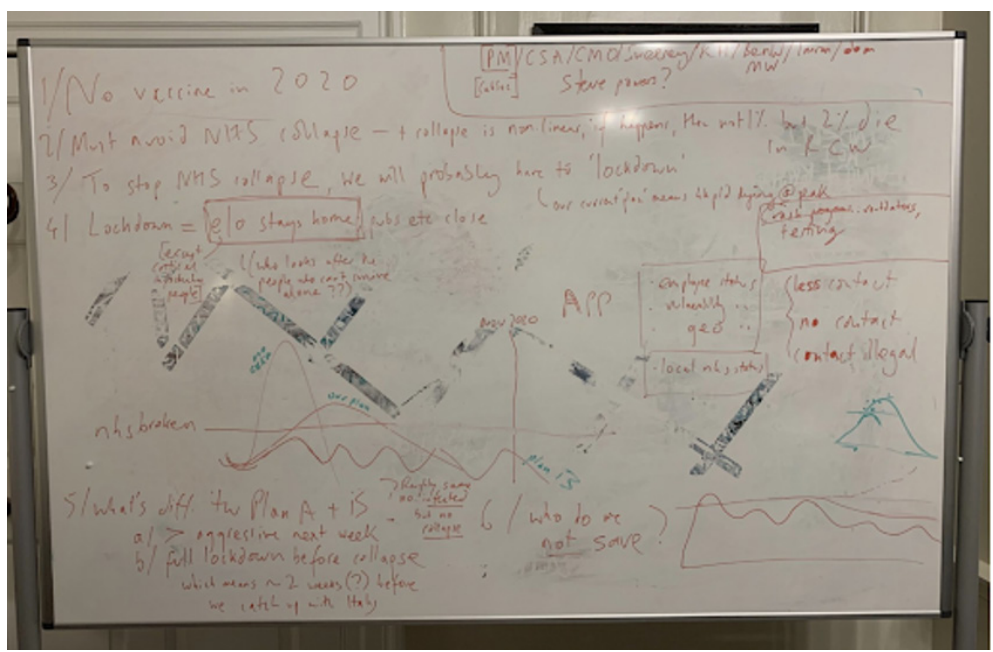

What this extreme case exhibits is a situation reflected in policy making more generally where vast amounts of data may be available, but there is a difficulty translating it into informed action (Luthfi and Janssen 2019). This can be down to the quantity and quality of digital information adding to the feeling that we have a separate "digital" reality. This can be seen in a reluctance to engage with information until something happens to suddenly "connect" it to the "real world". In this case the sudden realisation that "abstract" models predicted severe impacts to hospital admissions, at a scale which could no longer be ignored.

We argue that the reluctance to treat digital information as 'real', is perhaps due to how we have split the world in two: the grounded, trustworthy world of "the physical"; objects that can be touched, shaped by our hands and passed to other people, regulated by a scarcity which appeals to our feelings of simplicity, and knowing right from wrong. The other world is "the digital"; objects that we can only grasp indirectly, which come with concepts and structures of abundance, but are often working in the service of large multinational companies. These digital objects are generally understood as new, suspect, and untrustworthy.

Weaving breaks this false dichotomy in ways that make it possible to critique the digital infrastructure we inhabit more effectively. Threads are digital in precisely the same way voltages in our smartphones are digital — we combine these discrete elements into patterns we can use. This is not merely an analogy, but a tangible reality, which can for example be seen in how a woven artefact is a digital representation of its own making. The discrete structure of a weave can be replicated exactly as a digital signal sent in physical form via textile, exhibiting the same properties as a digital signal sent via radio waves. A woven textile can pass through long distances (of time) and be read perfectly when it reaches its destination.

**Figure 3:** Section of the Hallstatt tablet weaving circa 800-400 BC in our tablet weaving simulator, with code, tablet rotations and pattern shown.

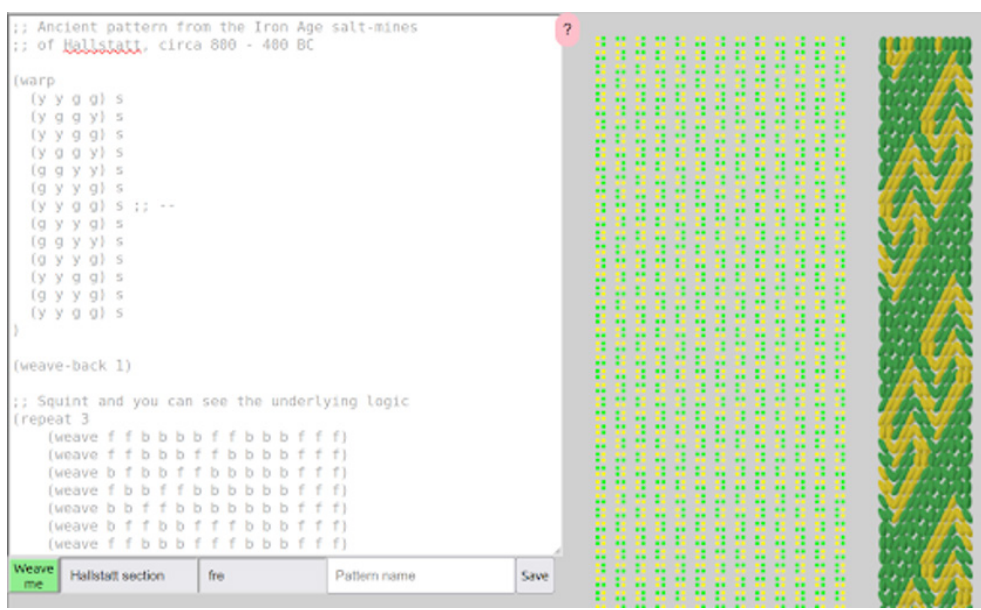

As an example of weaving signals travelling through time, archaeologists 'read' pieces of fabric such as the Hallstatt tablet weaving, which was discovered preserved in an Iron age salt mine. They are then able convert such data to an intermediate code that records the weaving tablet/card turning movements of the weaver from three thousand years ago. This code may then be followed to recreate the fabric, and fig 3. shows our own reconstruction through a custom simulator created by Dave Griffiths.[1] A reconstructed weave will be a perfect reconstruction, in terms of including all the micro-decisions (and indeed mistakes) of the original weaver. In this way, we can see that weaving looms are digital tools. They have passed through many more hands than the silicon-based digital tools we are more used to thinking about, and this perspective reveals how the human relationship with digital thinking goes back to prehistoric times.

As technologists, our interest in ancient weaving is therefore not an attempt to apply contemporary technology in understanding the past, but rather to apply heritage technologies in developing better understanding of the present. By grounding contemporary practice of live coding in understanding of heritage technology, we look to develop a healthier approach to contemporary programming languages technologies, that open up wider understanding of digital models and the impacts on our lives.

## 4. Revealing the Textility of Code

Discussions involving weaving and programming often incorrectly[2] refer to the development of Jacquard devices as the 'first computers.' There are, however, less well-told, more interesting connection points between textiles and contemporary computer technologies, such as Lisa Nakamura's work researching the involvement of the Navajo women who designed Fairchild semiconductors' first integrated circuits (Nakamura 2014). The company directly referenced the similarity of the traditional weaving designs to electronic junctions and transistors and used many images of Navajo designs in their company branding material.

---

1. The web-based tablet weaving simulator created by Dave Griffiths and used in the workshop is accessible at pattern-computer.thentrythis.org
2. All weaving, including at handlooms, can be computational. For example, see Lea Albaugh (2018) speaking on *"It's Just Matrix Multiplication": Notation for Weaving* for an explanation of the computational complexities of shaft looms aimed at computer programmers.

Is there some way that we can trace and unpick these threads of textile history in the central processing units of our devices today? Each processor has a defined set of instructions that it can execute — each instruction represented by a physical circuit of transistors built for the job. Therefore, the number of instructions needs to be minimised, as each one results in considerable expense. Something common to nearly all processors is that there are far fewer instructions which treat data as a numerical value (e.g., addition, subtraction, compare) compared to those that treat data as a pattern (shifts, rotates and other "bitwise" operations).

We can make these normally hidden operations visible, exposing the physical nature of the patterns that lie at the deepest levels of computation. As these processes are governed by the same rules as everything else (information is limited by physics; Shannon and Weaver 1963), it follows that there is no magical cyberspace, just combinations of voltages or threads we interpret as patterns with meaning.

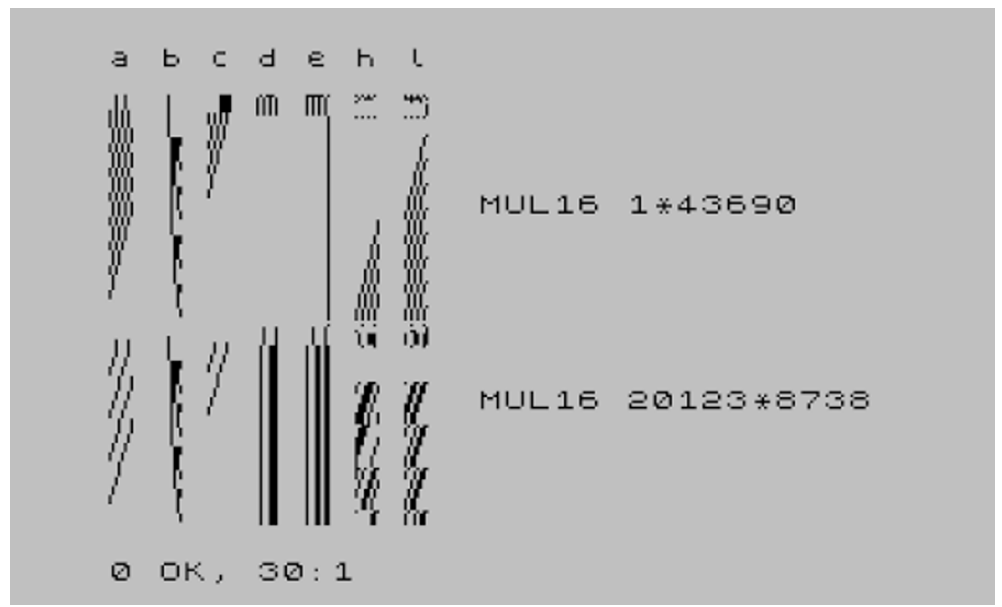

**Figure 4:** The Z80 processor comes from a similar era as Fairchild's integrated circuits and was a foundational design to modern microprocessors. Here we display the contents of its register memory after every microcode instruction of two multiplications of 16-bit numbers, revealing the patterns created.

During this workshop we followed this motivation of reconnecting live coders with textiles, by exploring replicating patterns from the Iron age (such as the Hallstatt textile) as well as Viking societies (see fig. 5 for examples). We previewed them in a simulation built for the workshop which included its own code representation of tablet weaving movements and tried out different variations to understand the logic of tablet weaving. The following section brings together some reflections from workshop participants.

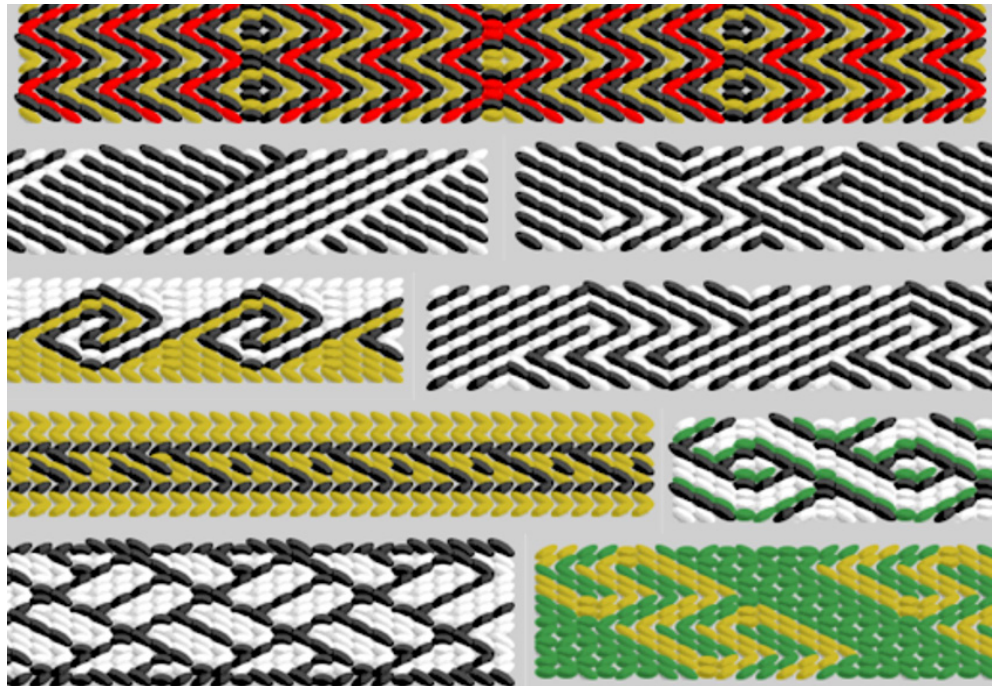

**Figure 5:** Selection of ancient Iron age and Viking tablet weaves.

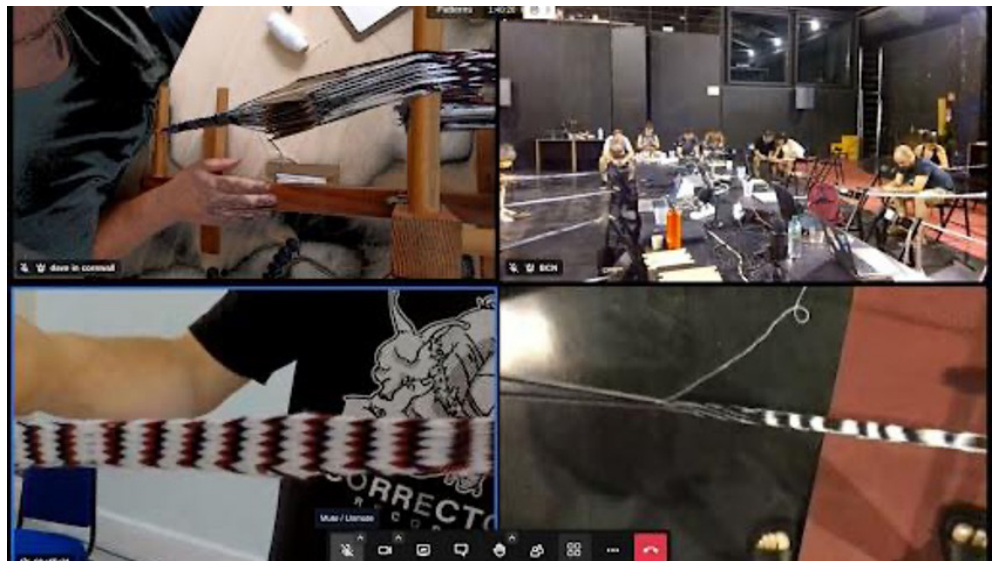

**Figure 6:** Zoom screenshot of the workshop.

## 5. Participant Reflections on the Workshop

Participants of the tablet weaving workshop each produced unique tablet weaving patterns, some of which are shown in Fig. 6. Although tablet weaving was the primary focus of the workshop, later, each participant was also asked to share their 'favourite pattern' with another person. Patterns could be of any nature, for example, sound, visual, etc., and could also include any of those reviewed in the workshop. Then, participants live-coded the other persons' patterns, exploring its possibilities. The results were briefly presented and shared, reflecting on why each one selected a particular pattern? What does it mean for them? And why do they consider their pattern a pattern?

283

The participants were surveyed four months after the workshop, gathering longer term reflections on the activity. The survey respondents represented below are Joana Chicau [JC], Flor de Fuego [FdF], Timo Hoogland [TH], Eloi Isern [EI], Michael-Jon Mizra [MJM], Iván Paz [IP], Roger Pibernat [RP], and Lizzie Wilson [LW]. We gave the following four prompts in the survey, to encourage the participants to reflect on the role of patterns in their work, and the relationship between code and textile.

The questions chosen to prompt the participants were chosen because of the interest in exploring different facets of live coders approaches, perspectives, and methodologies to approaching algorithmic pattern in the analogue practice of tablet weaving. These questions allowed them to explore their own relationship to coding practice and contrast this with the workshop activities. We also wanted to navigate the cognitive processes that a live coder might employ to approach this task, and whether they had existing strategies available to them that they could transfer to the practice of tablet weaving. Through these questions, the researchers also hoped to gain a deeper understanding of the relationship between coding practice and any other creative activities that relate.

(P1) How was your experience with the tablet weaving workshop? What was complex, what was simple? How does this compare or contrast from your experience with code?

(P2) The following is an excerpt from Joanne Armitage's paper "Spaces to Fail in: Negotiating Gender, Community and Technology in Algorave".

> For some, code emerges as a way of dealing with or organising life, for others code allows an expression of self, or a way of manipulating lived experiences and speaking back to them creatively. One person interviewed spoke about code as a way of working through their daily life, adding structures to it and providing functions for being. These lived patterns merge with their daydreams and expressions of colour and geometry to form her live coded visuals.

How does this relate to your life? Can you share an experience that compares or contrasts with it?

(P3) Do you enjoy a pattern-y craft or other pattern-y activity? E.g., weaving, braiding, origami, juggling, etc. If so, what does live coding and this activity give you that compares, and where do they diverge?

(P4) Have you thought about the workshop in the last few months, and if so what about it has stayed with you? Any influences on your thinking or makings?

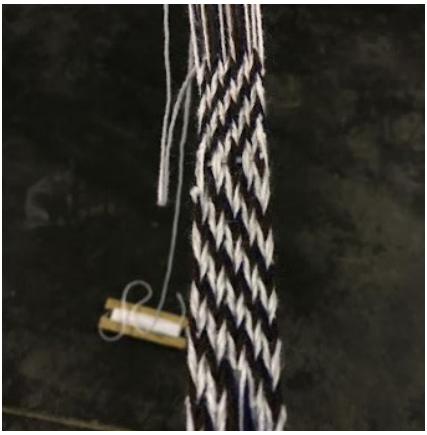
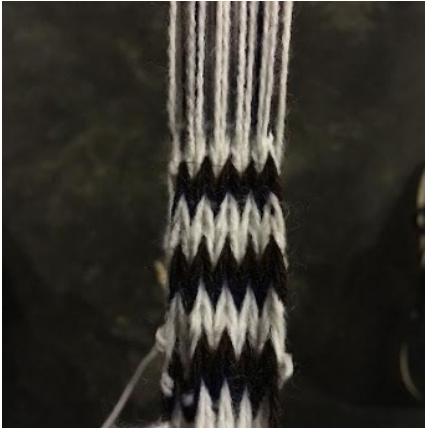
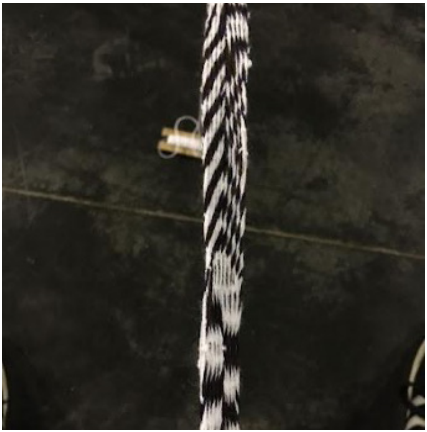
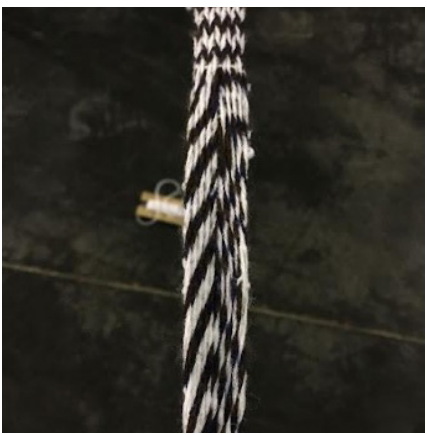

**Figure 7:** Emergent patterns from the participants' weave. Image credit: Timo Hoogland.

For the full set of participant responses, see the online repository for this paper, available at gitlab.com/algopattern/patterns-in-deep-time. Within those responses, we analysed the texts and found emerging themes from their answers about a variety of reflections that they shared.

### 5.1. Physicality

Perhaps the most apparent difference between these practices is that live coders generally work only with code, rather than directly with material as with tablet weaving. However, there is always more in the 'output' of live coding, whether music, choreography, or something else, than in the notation and rules for generating that output. This complexity contributed by material, and our perception of it, became particularly apparent when working with threads:

> I found more complexity in controlling the materiality of the "wool" than remembering the movements. The weaving algorithms were clear in my head, but knowing the right tension, the right pressure and where to stop pulling was difficult at the beginning. With the successive repetitions the movements felt more natural. [IP: P1].

In this answer the participant notes how the distinction between the cognitive processes and physical expression of them lead to some initial tensions for the participants. This was apparent for live coders, whose medium of expression — whilst still physical — relies heavily on cognitive processes. Live coding music has even been referred to as "the antithesis of physical musicianship" (Nilson 2007). Despite some initial struggles perhaps, a few of the participants noted the appeal of this "hands-on" approach at the workshop, where they became absorbed into the repetitive movements, making space for focussed creativity:

> I found the "hands-on" and movement focused character of weaving a smooth way of engaging in pattern making. For me muscle memory helps me a lot in making …over time, it became intuitive and fairly quickly I managed to improvise new patterns and explore more interesting combinations. [JC:P1]

> There was something quite enchanting about working with your hands and watching the patterns begin to appear. [LW:P1]

> I enjoy the fact that… making things in the moment, getting real time feedback from what you are making and not really being able to undo. [TH]

Philosophers, social theorists, and anthropologists have all spoken of the new reality that we inhabit in the twenty-first century due to the vast expansion of digital technologies, and that the digital era is incontestably new. However, viewed from another perspective, perhaps it can also be thought of as less of a colossal leap from the physical to digital eras. Humans have always had an urge to keep their hands busy, and this is perhaps one of the reasons heritage practices like weaving, spinning, and knitting were so culturally important. Typing on a keyboard, viewed through this lens, can be thought of as a natural progression of human behaviour. However, it is important to note that this does not reduce to the only reason why weaving may be culturally significant, especially when we note that textiles practices are often gendered as feminine.

The progression of materiality from human-material to human-machine embodies the demarcation of the physical to digital progression, but weaving exists as an intermediary, where the human is in close contact with both the fabric and the machine. In weaving, "bits" are manipulated in real-time whilst in coding the abstraction of bits is manipulated through language, and by extension typing on a keyboard. There are comparable abstractions in weaving, but these abstractions also take physical form, in the grouping of threads in shafts or tablets, the mapping of these groups through tie-ups and combination through treddling. By grounding live coders in this materiality, we hope this regression through human history allows them to make connections in how human-material loops and human-computer loops differ (e.g., in perceiving output and shifting behaviour).

### 5.2. Visualising Algorithms

A few of the participants reflected on how the workshop led them to contextualise algorithms in visual terms. Visualisation can be understood to leverage the visual system and augment human intelligence as a way to understand abstract processes (Engelbart 1962). Indeed, algorithmic practice has many connections with spatial processes or abstractions that might require a strong sense of cognitive visualisation process, and especially live coding languages with a more functional approach. For example, it often requires an understanding of ideas from geometry e.g., rotations, shifts, iterations; or linear algebra e.g. matrices or larger abstract structures and transition probabilities. One of the participants made the connection with how they use visualisation within mathematics, but drew a distinction between their experience of mathematical visualisation and what they were experiencing with the weaving:

> I loved the conscious experience of following an algorithm, understanding it to the point that I can almost predict the result of a small variation, this has offered me a different experience of visualizing the algorithms that I normally use in maths, as if the process that they described had a more material presence in the physical time and space. [IP]

The experience that the participant is noting is how visualisation which usually takes place as a cognitive process becomes a physical one. This became especially apparent when the relationship between the algorithm and physical space became unified. Other participants experienced this unification between the physical and cognitive worlds and found this became clearer as the workshop progressed:

> I feel that the process of abstracting a concept is a process of gradual reduction. To atomise in this way grants one the gift of microscopic analyses. I also think this expands outwards, with the generalising description of systems, which conversely implores one to analyse at the macroscopic level. I can also therefore relate to the experience of the artist who found inspiration in their daydreams about patterns - once one starts thinking in this way, the world seems to respond in kind. [MJM]

> I also liked the newness, the mapping between what I was doing physically and what was coming out of the weave wasn't always clear at first, but the more I navigated through the weave the more things started to become apparent. [FDF]

> ...Trying out different rotations of the tablets and repeating my randomly thought-of algorithms to see what the pattern is that emerges over time. In some ways it fits my approach to programming music and visuals, where I can have an idea of an algorithm I would like to explore, starting with the "what if...?" question, and then see what happens from there over time. [TH]

### 5.3. Satisfaction in Error

Error is a common, and oftentimes celebrated, feature of live coding performance. One viewpoint of error is the divergence of the observed output and its intended value. If we frame error in this way, it can also be a source for providing creative impetus, if the unexpected provides us with surprise and/or fulfilment. As it happens in live coding performance, where missing a coma or writing an extra digit is a frequently occurring failure (Knotts 2021; Roberts and Wakefield 2018), errors were present while weaving, maybe twisting in the "wrong" direction or not applying the "right tension". Errors contrast with what we had in mind, what we expect, and the results can surprise us in different ways. As in the practice of live coding, the live weaving action makes it easy to make mistakes, but those mistakes allow us to open up new avenues of exploration.

Some of the live coders who participated in the workshop compared the ways in which they encountered error in live coding with how they were experiencing it in the weaving workshop:

> For me the "trial-and-error" approach worked pretty well. When programming music, I can make an educated guess on what I can expect to happen, while with the tablet weaving this was not so much the case since I was completely new to it. This resulted in some interesting surprises of patterns that came out. [TH]

> As with coding, some complexity of the system began to arise when errors started to occur. It was relatively easy to undo sometimes, but there did seem times when small perturbations from what the instructor was doing felt that it shifted the outcome quite far. As with coding though, this did sometimes produce surprising and unexpected results that forced me to engage creatively with the weaving process. [LW]

> …'what is a pattern' is a question which positions itself between two poles; complexity/noise and simplicity/periodicity. These poles influence my approach to sound, where one can approach the construction of complex waveforms by the summation of simple wave forms, or one can construct wave forms through the use of stochastic processes. I am also intrigued by the human capacity to recognise patterns, and how we exploit our limited bandwidth to create pseudo-random functions. And this leads me to wonder about the nature of true randomness, whether it is obtainable, and what does it mean for these two poles to exist in a universe that is both probabilistic and deterministic. [MJM]

> The patterns that came out were really surprising, and it did feel a bit like getting unexpected results from code. [RP]

The notion of fulfilment from surprise is well researched within the context of the aesthetic experience of music. From a music-analytical standpoint, it has been argued that the creation and subsequent confirmation or violation of expectations is essential to aesthetic experience and the musical communication of emotion and meaning (Narmour 1990). Huron (2008) discusses what gives anticipation or surprise their distinctive phenomenological characters, and also how enforcing repetitions builds an expectation in the listener, and the subsequent violation of these expectations elicits a physical response. For others, they made note of what they found fulfilling in this task:

> The complex results, out of simple pattern-moves, were really satisfying. Maybe what I like about code is that it allows me to

> twist logic into poetic ways, which probably could be called a means of expression. [RP]

> I try to look for code as an expressive tool for communicating and connecting with different disciplines. [FdF]

> For someone like me, who does not come from a computer background, it was amazing to see the possibilities that opened up when working with the loom. [EI]

From these responses we can see that fulfilment not only came from the elements of error or surprise, but also from how the weaving allowed complex behaviour to occur and its associated poeticism, its wide-reaching capacity for expression and how the extensive possibility space it offered the weaver/live-coder can trigger new creative behaviours.

### 5.4. Patterns Passed On, Conveying Meanings

Patterns are polysemic, as are melodies or fabric patterns. They are read, felt, and interpreted in different ways. Sharing ideas, such as selecting a favourite pattern, among the participants visualises the different ways we interpret patterns by giving us perspective of others' experience.

> The exercise when people were asked to choose their favourite pattern and then pass it on to the next person to code in their own preferred language / software was interesting. That stayed with me, this idea of a collective string within which patterns are passed on 'hand-in-hand', reinterpreted and creating a lineage of patterns. [JC]

> I remember Dave saying something like this was a message that had thousands of years distance. And the idea of a weaving as a message, which I already somehow was aware because in Argentina we have that kind of idea with traditional weaving. [FdF]

> I really liked the accompanying computational representation that was going on, and tried to do a code representation myself to help try and parse what was happening... I also liked looking at the examples and seeing the way different cultures had their own representations of pattern that convey different meanings. It made me think about how music also conveys cultural meaning, and I wondered if there was any way of connecting these ideas of representation to musical representations (e.g., scores). [LW]

These responses suggest that the importance of conveying meaning in patterns lies not only in their aesthetic qualities, but also in

289

their ability to serve as expressive means of communication. Patterns play a crucial role in conveying meanings and cultural values. They often hold much historical and social significance, serving as a means of communication and self-expression for individuals and communities. Live coders, for example, recognise the importance of patterns in their community, many choosing to share their code and pass them on. Passing on patterns from person-to-person, or even generation-to-generation, is an essential way of conveying information and ideas, developing cultural heritage and even promoting intergenerational continuity. By recognizing and valuing the many ways in which patterns can be interpreted and experienced, we can deepen our appreciation for the richness and complexity of human culture and creativity.

## 6. Conclusion

Tablet weaving is an action through which algorithmic processes materialise. As in code, unanticipated results happen through chaotic interaction, and these surprises can be creatively explored. These unplanned experiences, especially present when trying something new, are intrinsic to both weaving and coding. Weaving connects the physical materiality of the woven patterns with their immaterial algorithmic descriptions. It allows us to visualise the algorithmic processes that describe the instructions from which they emerge. Weaving together in a group, following and deviating from the instructions we were given, gave us shared perspective when comparing our experiences. The central discussions on the workshop, as it was attended mainly by live coders, revolved around algorithms, time, error, repetition, and codification, but also about the way we interpret patterns such as rhythm, regularity, and how easy it is for us to recognize or perceive a pattern, i.e. the limits of our spatial and temporal perception, and of our ability to predict the outcome of the algorithms we create and modify. It is interesting that these ideas (descriptions) match the ways in which we describe the material and immaterial aspects of patterns. The examples included in the simulator, ranging from prehistoric Iron age to Viking, added an extra layer to the different dimensions of the patterns: the way different cultures had their own representations of patterns (sometimes closely related) that convey different meanings.

Our half-day workshop created a scene of digital artists used to working with computers, finding themselves working with the twisting and interlacing of threads, while still thinking about code. Perhaps this is a vision of the future rather than a re-enactment of the past. With environmental breakdown and health emergencies triggering a growth of 'collapse computing' culture, coders may need to reconnect with numerical crafts such as hand-weaving, to maintain their interests in digital art. More optimistically, this re-uniting of coding and textile culture could lead to a richer, more sustainable, tangible, and culturally-grounded approach to future technology.

**Acknowledgements.** This work was part-funded by UKRI Future Leaders Fellowship [grant number MR/V025260/1], and by Creative Europe via the on-the-fly project. Additionally, Wilson's contributions were supported by EPSRC and AHRC under the EP/L01632X/1 (Centre for Doctoral Training in Media and Arts Technology)